\documentclass[%
 aip,
rsi,%
 amsmath,amssymb,
 reprint,%
floatfix
]{revtex4-1}
\usepackage{xcolor}
\usepackage{graphicx}
\usepackage{dcolumn}
\usepackage{bm}
\usepackage{color}
\usepackage{fancyhdr}

\begin{document}

\preprint{AIP/123-QED}
\thispagestyle{fancy}
\title{A Fully Automated Dual-Tip SNOM for Localized Optical Excitation and Detection in the Visible and Near-Infrared}

\author{Najmeh Abbasirad}%
\email{najmeh.abbasirad@uni-jena.de}
\affiliation{Institute of Applied Physics, Abbe Center of Photonics, Friedrich Schiller University Jena\\Albert Einstein Str.6, 07745 Jena, Germany}%
\author{Jonas Berzins}%
\affiliation{Institute of Applied Physics, Abbe Center of Photonics, Friedrich Schiller University Jena\\Albert Einstein
Str.6, 07745 Jena, Germany}%
\affiliation{The Netherlands Organization for Applied Scientific Research, TNO, 2628CK Delft, The Netherlands}
\author{Kenneth Kollin}
\affiliation{RHK Technology, Troy, Michigan, USA}
\author{Sina Saravi}
\affiliation{Institute of Applied Physics, Abbe Center of Photonics, Friedrich Schiller University Jena\\Albert Einstein Str.6, 07745 Jena, Germany}%
\author{Norik Janunts}
\affiliation{Leibniz Institute of Photonic Technology\\Albert Einstein Str. 9, 07745, Jena, Germany}
\author{Frank Setzpfandt}
\affiliation{Institute of Applied Physics, Abbe Center of Photonics, Friedrich Schiller University Jena\\Albert Einstein Str.6, 07745 Jena, Germany}%
\author{Thomas Pertsch}
\affiliation{Institute of Applied Physics, Abbe Center of Photonics, Friedrich Schiller University Jena\\Albert Einstein Str.6, 07745 Jena, Germany}%
\affiliation{Fraunhofer Institute for Applied Optics and Precision Engineering\\Albert Einstein Str.7, 07745 Jena, Germany}%

\date{26 April 2019}

\begin{abstract}
Near-field optical microscopes with two independent tips for simultaneous excitation and detection can be essential tools for studying localized optical phenomena on the subwavelength scale. Here we report on the implementation of a fully automated and robust dual-tip scanning near-field optical microscope (SNOM), in which the excitation tip is stationary while the detection tip automatically scans the surrounding area. To monitor and control the distance between the two probes, mechanical interactions due to shear forces are used.
We experimentally investigate suitable scan parameters and find that the automated dual-tip SNOM can operate stably for a wide range of parameters. To demonstrate the potential of the automated dual-tip SNOM, we characterize the propagation of surface plasmon polaritons on a gold film for visible and near-infrared wavelengths. The good agreement of the measurements with numerical simulations verifies the capability of the dual-tip SNOM for the near-field characterization of localized optical phenomena.

\end{abstract}


\newpage
\thispagestyle{fancy}
\maketitle
\tableofcontents

\section{Introduction}
Scanning near-field optical microscopy (SNOM) is an invaluable tool to investigate optical phenomena within the near-field region of nano-photonic structures. Since its first realization, \cite{Lewis1984,Pohl1984} SNOM has been a measurement tool in various research areas such as plasmonics\cite{Kawata2009,bazylewski2017review}, photonic crystals \cite{Rotenberg2014,spasenovic2012measuring,huisman2012measurement}, surface-enhanced Raman spectroscopy\cite{zhang2016near,Kusch2017} and photovoltaics\cite{Xiao2017,Mensi2018}.
The tip in a conventional aperture SNOM often serves for either near-field excitation or detection, which can be more localized than allowed by the diffraction limit. However, in the case of a near-field excitation by the SNOM tip, the measurement involves a diffraction limited detection in the far field, which cannot achieve a spatial resolution below the diffraction limit. Similarly, when the SNOM tip is utilized for near-field detection, the far-field excitation is diffraction limited. A dual-tip SNOM which uses two independent aperture tips can enable simultaneous excitation and detection below the diffraction limit, where the tips can be positioned precisely with respect to each other. Such a capability in spatial positioning  of a point-like emitter and detector allows mapping of the Green's function for any photonic structure.
\\Dual-tip scanning probe microscopes (SPMs), such as scanning tunneling microscope and atomic force microscope (AFM), have been already established as indispensable instruments for nano material characterization and manipulation\cite{nakayama2016multiple,Voigtlander2018}. Although a single-tip SNOM has been utilized in combination with other scanning tip microscopes, e.g. atomic force and scanning electron microscope (SEM)\cite{Haegel2014}, only a few works regarding dual-tip SNOM  measurements have been reported, all of which were performed within the visible spectrum \cite{Dallapiccola2009,ren2011interference,Liu2014,klein2014polarization,klein2017dual}. In these works scanning the whole area around the excitation tip was not feasible due to the collision of the tips. Therefore, the scan area was either chosen sufficiently far from the excitation tip\cite{ren2011interference} or the detection tip was stopped manually to avoid the collision\cite{klein2014polarization}.
\\
Due to the growing interest in tailoring light by means of plasmonic or dielectric metasurfaces\cite{meinzer2014plasmonic,staude2017metamaterial}, the availability of a reliable instrument to investigate the localized optical response within the near-field region becomes paramount. Moreover, the distinguished capability of the automated dual-tip SNOM in near-field excitation and detection would provide new possibilities for the investigation of the localized near-field response, e.g. for disordered photonic systems\cite{Caselli2017,garcia2017physics} where the response sensitively depends on both the exact position of the excitation and the exact position of the detector.
In an ideal dual-tip SNOM, the excitation tip is stationary while the detection tip automatically scans the whole area surrounding the excitation tip to detect the accessible near-field information. Since there is a desire to detect the near-field optical response in close proximity of the excitation point, the main challenge in the realization of the dual-tip SNOM is a reliable and robust technique to prevent the collision and damage of the excitation and detection tips when they are close to each other. It has been demonstrated that mechanical interaction between two oscillating tips can serve as a reliable indicator of the distance between the two tips\cite{Klein2012}. 
Such mechanical interaction is induced by shear forces between the tips and leads to a perturbation of the oscillation of one tip by the oscillating motion of the other tip. The induced change in the mechanical response of the tips is detected and used to monitor their distance.
In the first demonstration of collision avoidance in a dual-tip SNOM the authors used two tips oscillating parallel to the sample plane with identical resonance frequencies.\cite{Kaneta2010,Kaneta2012,Fujimoto2012}This extra oscillation is orthogonal to the direction of the detection tip's resonant oscillation but in the same plane, to allow frequency demodulation by a heterodyne technique. However, the extra oscillation amplitude gives rise to an additional motion, which in turn limits the image resolution. Moreover, the low-frequency demodulation is more susceptible to the electronic and vibrational noises of the system, which then negatively effects the distance control between the two tips, and hence the overall scanning stability.
In our work, we demonstrate an automated dual-tip SNOM technique, which resolves these shortcomings. We use two tips oscillating perpendicular to the sample plane, with two different resonance frequencies, eliminating the need for the extra in-plane oscillation and its detrimental effect on the scanning resolution. Moreover, to implement our distance detection scheme we use the high resonance frequency of one tip's oscillation to demodulate the signal coming from the other tip when they mechanically interact. This leads to an increased signal-to-noise ratio (SNR) in the measurements, resulting in a stable distance control between the two tips.
Additionally, we use a fully digital controller to improve the overall SNR of the system. The controller simultaneously controls the two tip electronics and immediately converts analog signals to digital signals, carrying out all the analysis in the digital domain, which significantly lowers the noise of the system.

\begin{figure}
\includegraphics[width=8cm, height=3.5cm]{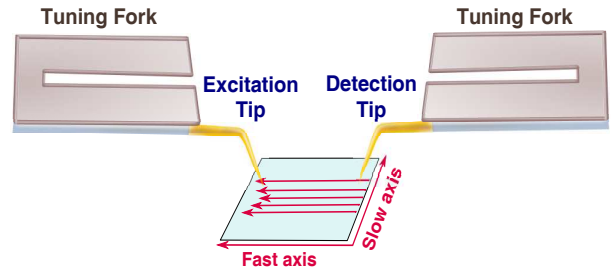}
\caption{Scheme of the tips and scan configuration of the dual-tip SNOM. Red arrowed lines demonstrate the single direction of the scan in the automated dual-probe SNOM setup. The lines vary in length due to the presence of the excitation tip.}
\label{twotips}
\end{figure}
This paper presents the implementation of this fully automated dual-tip SNOM and experimentally verifies its performance. Furthermore, to demonstrate the potential of the dual-tip SNOM, we have measured the near fields of surface plasmon polaritons (SPPs) propagating on a rough gold film at visible and, for the first time, at near-infrared excitation wavelengths.

\section{Implementation of the fully automated dual-tip SNOM }
To realize a robust automated dual-tip SNOM, we have integrated two SNOM heads (MV-4000, Nanonics Imaging Ltd) with a fully digital SPM controller (R9, RHK Technology, Troy, Michigan, USA). In the dual-tip SNOM measurements we have used commercially fabricated aperture tips (Nanonics Imaging Ltd), which are tapered-end optical fibers with a 200-300 nm thick metal coating. The aperture tips are tilted by a 30-degrees angle with respect to the sample normal to allow the combination of the two tips as shown schematically in Fig. \ref{twotips}. Each cantilevered aperture tip, attached to a tuning fork, is mounted on the SNOM head. The SNOM heads consist of a stepper motor for coarse adjustment of the tip position on the sample and a piezoelectric scanner for the fine movement and adjustment of the tip height over the sample during the scan. Fig. \ref{twotips} schematically presents our dual-tip SNOM configuration in which the excitation tip is stationary while the detection tip raster scans the surrounding area of the excitation tip. The red-arrowed lines demonstrate a single scan direction in automated dual-tip SNOM. The scan lines along the fast axis become shorter due to the presence of the excitation tip and the geometrical shape of the two tips. Thus, there is a parabolic region that is not accessible due to the excitation tip and is excluded from the scan area.

\begin{figure*}
\includegraphics[width=14cm, height=8cm]{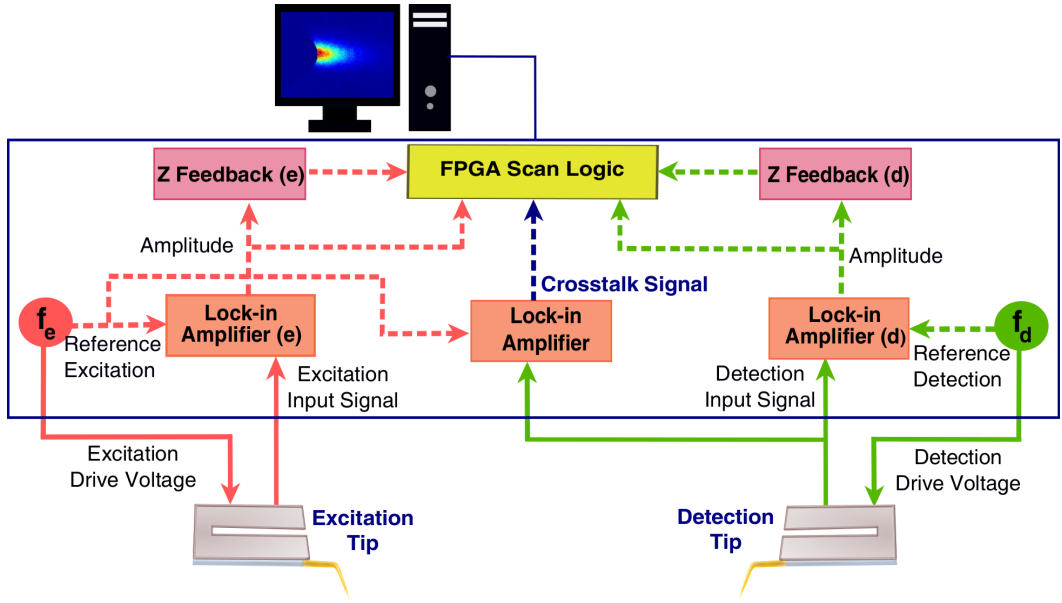}
\caption{Scheme of the control electronics of the automated dual-tip SNOM. The signal from the detection tip is associated with green color and red color is linked to the signal from the excitation tip. Dashed lines represent purely digital signals.}
\label{setup}
\end{figure*}
Both tips of our dual-tip SNOM, connected to the tuning forks, operate in a tapping mode, where they oscillate vertically. To induce the oscillation, the piezoelectric tuning forks are driven by sinusoidal voltages at their fundamental resonance frequencies. The induced mechanical oscillation leads to oscillating charges which are measured as an electric current. The transimpedance amplifier converts the current to a voltage which is digitized by the controller before being demodulated by an internally integrated lock-in amplifier. Throughout the paper, we refer to the amplitude of this voltage as the oscillation amplitude, which describes the strength of each tips' oscillation. The excitation tip oscillates at the resonance frequency $f_e$, whereas the detection tip oscillates with $f_d$. Different to the automatic dual-tip SNOM implementation reported earlier\cite{Kaneta2012}, our instrument uses tuning forks with two different resonance frequencies for the excitation and detection tips. These resonance frequencies are the fundamental resonance frequency of each tip which could vary between 30-40 kHz. In order to enable discrimination between signals coming from the two tips, the difference between the two resonance frequencies should be larger than the frequency resolution of the used lock-in amplifier. In our measurements, the difference between the resonance frequencies of the tips is usually about 3-5 kHz. In the tapping mode, the extension of the tip oscillation is large compared to the range of effective forces originating from the sample surface \cite{voigtlander2016scanning}, however, the interaction with the sample perturbs the motion of the tips. This perturbation of the mechanical motion is again transformed to an electrical signal by the tuning forks and monitored by dedicated lock-in amplifiers which are locked to $f_e$ and $f_d$, as shown schematically in Fig.~\ref{setup}. The Z feedback controls the tip approach to the sample and regulates the tip-sample distance during the scan. A field-programmable gate array (FPGA) processes simultaneously the digital data from different components inside the controller to improve the response time of the system when the two tips interact. 
Fig.~\ref{setup} illustrates the schematic of the automated dual-tip SNOM and its electronics.
In order to detect mechanical interaction between the tips induced by shear-forces, another lock-in amplifier is required, which demodulates the oscillation signal of the detection tip at the resonance frequency of the excitation tip. The output of this lock-in amplifier is called crosstalk signal. For well separated tips, no mechanical interaction takes place. The detection tip oscillates only with the frequency $f_d$ and the crosstalk signal is at the noise level of the electronic system which we refer to as a base crosstalk signal. When the detection tip approaches the excitation tip during the scan, the two tips interact and mechanically couple due to the shear-forces. Now the response of the detection tip will contain signals with the resonance frequency of the oscillation of the excitation tip and the crosstalk signal significantly increases. Hence, the crosstalk signal can be employed as a reliable proximity indicator of the tips.
\\To automatically implement the collision prevention scheme, we have developed a dual-tip detection function in the software of the controller. A predefined threshold value for the crosstalk signal, called avoidance threshold, should be set by the user in this function. While scanning, the detection tip moves on a line along the fast axis (see Fig.~\ref{twotips}), approaching the excitation tip. Once the crosstalk signal reaches the avoidance threshold, the detection tip stops scanning, sweeps back and continues scanning the next line. By repeating this procedure for each scan line, the image of the whole scan area is mapped and the avoidance area builds up as a parabolic region within the scan area. It is worth mentioning that the well-defined parabolic region is an indication of the stability of our distance control scheme.
The thickness of the coatings of the tips and the aperture diameter determine the size of the tip apex. Therefore, the minimum distance between the centers of two apertures is usually between 0.5-1 $\mu$m. This implies, that the optical signal can only be mapped by the detection tip at distances that are larger than this minimum distance from the point of excitation.

\section{Stability evaluation of the dual-tip SNOM}
To evaluate the stability of the dual-tip SNOM measurement we first performed topography measurements on a flat sample. To this end, we used a silicon wafer with an almost defect-free surface. Therefore, any significant variation in the measured height can be attributed to instabilities of the automated dual-tip SNOM. We evaluated the stability of the automated dual-tip measurement for two measurement parameters, the oscillation amplitude of the excitation tip at its resonance frequency and the avoidance threshold value.
We restricted our analysis to these two parameters, as they are specific to the dual-tip SNOM measurement. 
\\
The influence of the parameters for single-tip SNOM measurements such as the oscillation amplitude of the moving detection tip or the settings of lock-in amplifiers and feedback loops were optimized in our system and kept constant during all measurements reported here. The amplitude of the detection tip was adapted to the sample characteristics to avoid tip or sample damage. In our case, we chose an oscillation amplitude equivalent to 280~mV. 
\begin{figure}
\includegraphics{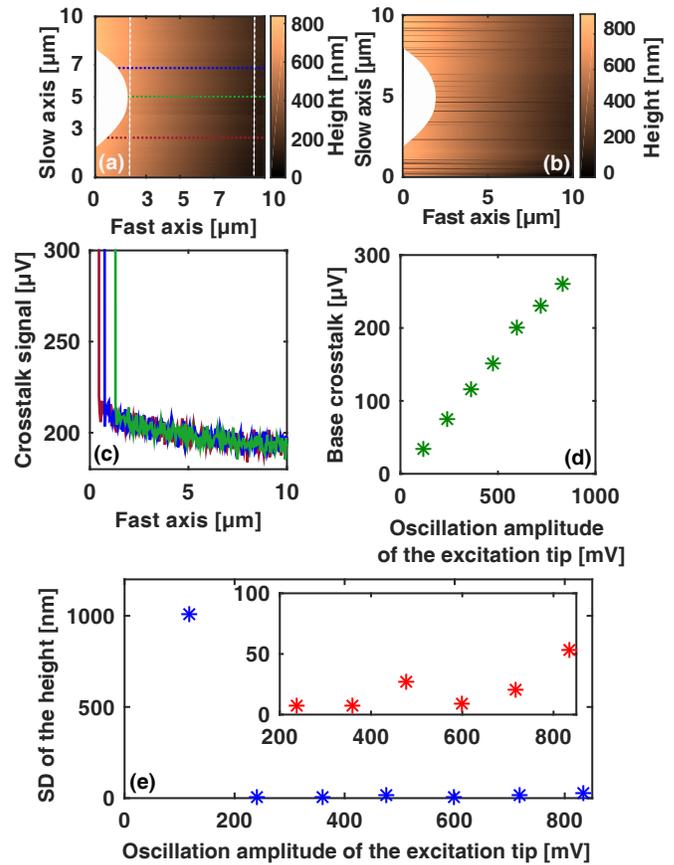}
\caption{(a) Raw height data of a silicon wafer measured with scanning detection tip when the oscillation amplitude of the excitation tip is 600~mV, (b) 840~mV. The white parabolic region is the avoidance area. (c) Crosstalk signals corresponding to the three topography lines denoted as colored dotted lines in (a). (d) Base crosstalk signal as a function of the oscillation amplitude of the excitation tip. (e) Standard deviation (SD) of the height as a function of different oscillation amplitudes of the excitation tip while the oscillation amplitude of the scanning detection tip is equivalent to 280 mV in all topography measurements. The standard deviation of the height was calculated for the corrected height in the same area between the white dashed lines in (a). The inset shows the zoomed in region of the data for the oscillation amplitudes of the excitation tip between 215-840 mV.} 
\label{SDRA}
\end{figure}
The excitation tip is stationary and does not scan the sample. Hence, there is more freedom to choose among different oscillation amplitudes. The resonance frequency of the excitation tip was $f_e=34.96$ kHz and the detection tip $f_d=40.12$ kHz.
\\ To investigate the effect of the oscillation amplitude of the excitation tip on the measurement's stability, we map the topography of 10x10 $\mu{m}^2$ of the flat Si surface for different values of this parameter.
We have selected a maximal oscillation amplitude of 840~mV for the excitation tip, which ensures that tip and sample are not damaged.
Furthermore, in this set of measurements we have fixed the avoidance threshold value. Changing the oscillation amplitude of the tips affects the base crosstalk signal appearing from an intrinsic electronic noise even when the two tips are far separated. We have chosen the avoidance threshold based on the maximal oscillation amplitude of the excitation tip at 840~mV. In this case, the measured base crosstalk signal was 280~$\mu$V. Therefore, the avoidance threshold was set to 300~$\mu$V, 20 $\mu$V larger than the base crosstalk signal.  For any value of the avoidance threshold between 280-300 $\mu$V the detection tip does not move because the crosstalk signal meet the avoidance threshold due to an intrinsic noise of the system. To study the effect of the amplitude of the excitation tip, we performed dual-tip SNOM measurements between 840~mV of oscillation amplitude and the minimum value of 120~mV at which the measurement is not stable anymore due to the tips frequent collision. 
The results of these measurements are summarized in Fig.~\ref{SDRA}. The raw data of the topography measurement when the oscillation amplitude of the excitation tip was 600~mV and 840~mV are demonstrated in Fig.~\ref{SDRA}(a) and (b), respectively. The measurement shown in Fig.~\ref{SDRA}(a) was stable. The linear increase in the height is associated with the tilt of the scanner plane with respect to the sample. On the other hand, Fig.~\ref{SDRA}(b) indicate an unstable measurement, where the instabilities show up as an artifact in the raw data of the height.\\
The crosstalk signal is always recorded simultaneously along with the topography line data. We selected three topography lines in Fig.~\ref{SDRA}(a) to plot the corresponding corsstalk signals. In Fig.~\ref{SDRA}(c) the respective crosstalk signal of a selected topography line is denoted by the same color. The crosstalk signal slowly increases as the two tips approach each other, which is due to the weak mechanical coupling. The sudden rise in the crosstalk signal only occurs when the two tips are within the effective range of the shear forces. Therefore, the crosstalk signal exceeds the avoidance threshold. The last topography data in each line is obtained at the pixel in which the crosstalk signal exceeds the threshold value.
\\ As a cumulative measure for the scan stability we calculate the standard deviation of the corrected height in the area marked by the dashed lines in Fig.~\ref{SDRA}(a). To obtain the corrected height the measured raw data was corrected for a tilt of the scanner plane with respect to the sample as well as the scanner bow by subtracting a fitted paraboloid. The standard deviation $\sigma$ of the corrected height is calculated by the following formula,
\begin{equation}
\sigma=\sqrt{\frac{\sum_{x=1}^{M}\sum_{y=1}^{N}{(h(x,y)-\overline{h})}^2}{MN}.}
\label{equ1}
\end{equation}
Here, $h$ is the corrected height and $\overline{h}$ the average corrected height. $M$ and $N$ are the number of pixels along the width and length of the area within the dashed lines in Fig.~\ref{SDRA}(a). The base crosstalk signal and the standard deviation of the height  as a function of the oscillation amplitude of the excitation tip are depicted in Fig.~\ref{SDRA} (d) and (e), respectively. 
The large value of the standard deviation of the height at an amplitude of 120~mV is due to the fact that here the base crosstalk signal is only 35~$\mu$V, which is much smaller than the avoidance threshold value of 300 $\mu$V. Hence, before the crosstalk signal meets the avoidance threshold, the two tips often collide during the scan and lose contact with the sample.
When the oscillation amplitude of the excitation tip is equivalent to 840~mV, there is another effect that leads to the artifact visible as lines in the topography measurement, which is observed in Fig.~\ref{SDRA}(b). In this case, due to the small difference, the crosstalk signal reaches the avoidance threshold
fictitiously due to inherent noise of the electronics, not actual closeness and interaction between the tips. As a result, the detection tip aborts the scanning of the current scan line before reaching the effective shear-forces region. The inset of Fig.~\ref{SDRA}(e) shows a zoom in of the standard deviation for the oscillation amplitudes of the excitation tip between 215-840 mV. Apart from the oscillation amplitude of 840~mV, in this parameter range we have not observed any instability during the dual-tip SNOM topography measurements. We assume that the small differences in the standard deviation of the height are due to fluctuations in the feedback loop during the measurements. We conclude that the measurements are robust for a wide range of oscillation amplitudes of the excitation tip. 
\begin{figure}
\includegraphics{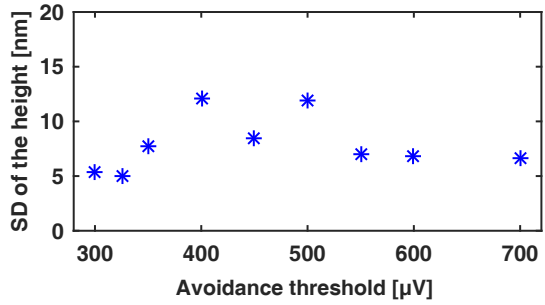}
\caption{Standard deviation of the height calculated for the topography measurements data at different avoidance thresholds, where the oscillation amplitude of the excitaion and detection tips are kept fixed to 600 mV and 280 mV, respectively.}
\label{SDAT}
\end{figure}
So far we have demonstrated that changing the oscillation amplitude of the excitation tip leads to different base crosstalk signals and its difference from the avoidance threshold determines the robustness of the system.
In another set of measurements to test the stability of our system, we kept the base crosstalk signal constant and the avoidance threshold was varied. For these measurements, we selected the oscillation amplitude of 600~mV for the excitation tip and, as before, 280 mV for the detection tip to ensure robust dual-tip SNOM measurements. These values of the oscillation amplitudes of the excitation and detection tips lead to a base crosstalk signal of 205~$\mu$V as was shown in Fig.~\ref{SDRA}(d). To avoid artifacts due to a small difference between the base crosstalk signal and the avoidance threshold, the minimal avoidance threshold was set to 300 $\mu$V. For avoidance threshold larger than 700 $\mu$V the crosstalk signal reaches the value of the avoidance threshold only after the collision of the tips. Such collisions lead to the tips losing their contact with the sample and ceasing the scanning which is not desirable. In Fig.~\ref{SDAT}, we have plotted the standard deviation of the corrected height as a function of avoidance threshold between these limiting values. For all values between 300-700 $\mu$V, stable dual-tip SNOM measurements are possible. We again attribute the slight variation of the standard deviation of the height between 10-20 nm to fluctuations of the feedback loop during the scanning.

\section{Near-field excitation and detection of surface plasmon polaritons}
\begin{figure*}
\includegraphics{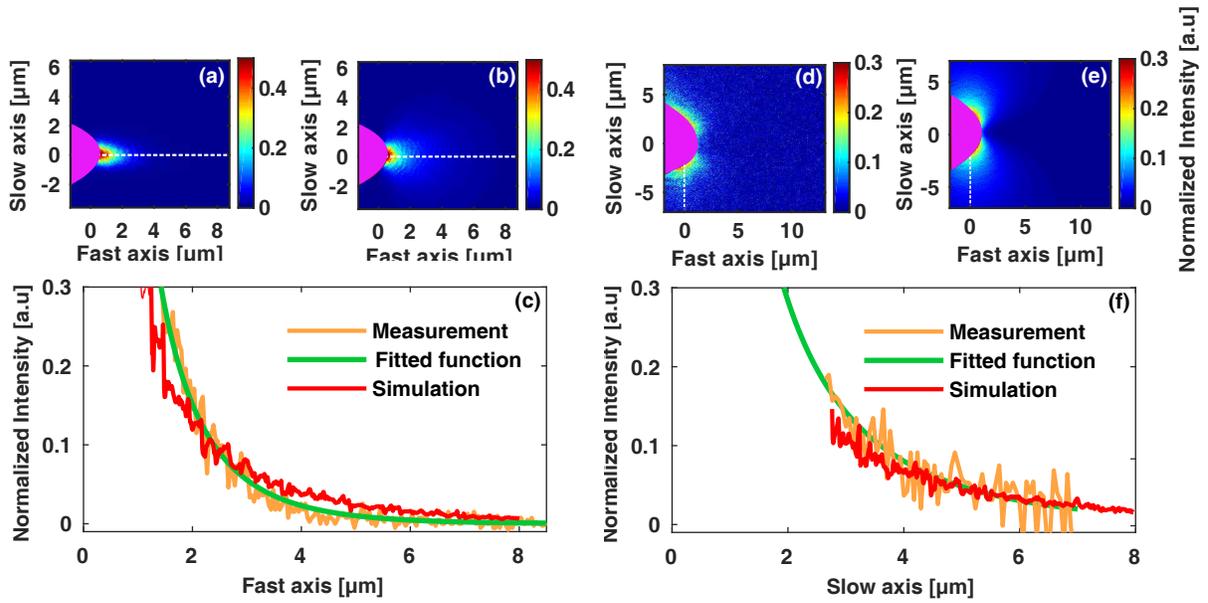}
\caption{(a) Measured radiation pattern of the SPPs on gold film at a wavelength of 633 nm. The zero points along the slow and fast axes represent the center of the excitation tip. The pink parabolic region represents the avoidance area. Color scales are normaized intensity with an arbitrary unit. (b) Simulated intensity of the transverse components of the magnetic field in the plane of the gold film. The white dashed lines in (a,b) pass through the symmetry axes of the SPPs radiation patterns. (c) Plot of the intensity line profiles corresponding to the dashed lines in (a,b). The measured line profile was fitted to the function  $Ae^{-r/l_{sp}}/{r}$, which is also shown. (d) Measured SPPs intensity at a wavelength of 1550 nm and (e) the corresponding simulation result for the transverse magnetic field. (f) Plot of the intensity line profiles corresponding to the dashed lines in (d,e) along with the function $Ae^{-r/l_{sp}}/{r}$ fitted to the measured data in (d).}
\label{visible_infra}
\end{figure*}
To show the capabilities of the automated dual-tip SNOM, we have measured the propagation of SPPs on gold thin film with a rough surface.
To carry out the automated dual-tip SNOM measurements, a 200~nm thick gold film  was used. The film has a root mean squared (RMS) roughness of 8 nm, measured by an AFM. To excite SPPs in the visible spectral range, the gold film was excited by a single-mode aperture fiber tip with 148 nm diameter at 633 nm wavelength. Such localized excitation produces a dipole-like emission pattern for the generated SPPs \cite{Hecht1996}.
We have used a single-mode aperture fiber with 160~nm aperture diameter to scan the surface and detect the emission pattern of the dipole on the rough gold film.
Fig. \ref{visible_infra}(a) presents the mapped normalized intensity of the SPPs at wavelength 633 nm, which propagates along the fast axis of the scan range due to excitation with a magnetic field polarized along the slow axis. The acquisition time of the optical signal was set to 5 ms/pixel. The dark counts of the detector (PerkinElmer SPCM-AQR) were subtracted from the measured intensity before normalization. The zero coordinates of the fast and slow axes indicate the center of the excitation aperture, which is estimated based on the aperture size and the width of the coating at the apex using SEM image. The pink parabolic region on the left is the avoidance area, which restricts the mapped intensity to only one lobe of the radiation pattern. Fig.~\ref{visible_infra}(b) shows the result of numerically simulated SPPs intensity at 633 nm on a gold film by incorporating the measured RMS roughness as in the experiment. The numerical simulations were carried out by using commercial software based on the finite-difference time-domain method (Lumerical Ltd.), considering the excitation as a magnetic dipole\cite{klein2014polarization} polarized along the slow axis with a distance of 10~nm above the sample. Considering previous work, which had shown that the detection tip is more sensitive to the magnetic field, \cite{klein2014polarization} in Fig.~\ref{visible_infra}(b) we have plotted the simulated intensity of the transverse components of the magnetic field, which agrees well with the measurements. For the sake of intuitive comparison with the measurements, the avoidance area in the simulation is covered. To compare the measurement with the simulation more precisely, the intensity line profiles corresponding to the white dashed lines in (a) and (b) have been plotted in Fig.~\ref{visible_infra}(c). These white dashed lines denote the symmetry axes of the SPPs radiation pattern. The measurement data was fitted to the function $Ae^{-r/l_{sp}}/{r}$ in which $l_{sp}$ is the propagation length and $r$ is the distance from the center of the aperture \cite{Hecht1996}. A propagation length of $l_{sp}=1.5$ $\mu$m was calculated from the fit function, which due to the roughness is much shorter than expected for ideal gold films\cite{Kolomenski2009}.
Fig.~\ref{visible_infra}(d) demonstrates a similar measurement at the wavelength 1550 nm  on the same gold film. To our knowledge this is the first demonstration of a dual-tip SNOM measurement at infrared wavelength. The excitation tip was a single-mode aperture fiber tip with 300~nm aperture diameter and the detection tip was a 500~nm multi-mode aperture fiber. Once more, the average dark counts of the near-infrared detector (id210, ID Quantique) were subtracted from the mapped intensity before normalization. This time, the input polarization along the fast axis leads to SPPs radiating along the slow axis. Thus, both lobes of the SPP radiation pattern could be measured. The missing central part of the two lobes in this figure is due to the avoidance region. Fig.~\ref{visible_infra}(e) shows the result of the corresponding numerical simulation of the SPPs radiation pattern, where the magnetic dipole excitation is oriented along the fast axis. Again, the intensity of the transverse components of the magnetic fields is plotted.
The intensity line profiles along the symmetry axis of the radiation pattern, given by the dashed lines in Fig.~\ref{visible_infra}(d,e) have been plotted in Fig.~\ref{visible_infra}(f) along one lobe of the radiation pattern. To calculate the respective propagation length the same function as before was fitted to the measured intensity line profile. In this case, a propagation length of $l_{sp}=3.5$ $\mu$m was calculated, again shortened by the surface roughness of the sample.  

\section{Summary}
We have developed a fully automated and robust dual-tip SNOM to measure the localized near-field response of nanophotonic structures. A reliable collision prevention technique has been realized based on the shear-force interaction between the two tips, which oscillate at different resonance frequencies. This leads to a very stable system which can operate in a wide range of the scan parameters, which we have demonstrated experimentally. Finally, the near-field characterization of propagating surface plasmon polaritons at visible and near-infrared wavelengths was demonstrated experimentally with sub-wavelength excitation and detection. These results establish dual-tip SNOM as a fully functional tool for the investigation of optical nanostructures.

\section*{Acknowledgement}
N.A. was funded by the German Academic Exchange Service (DAAD) via 
the Graduate School Scholarship Programm (GSSP) and the German Research
Foundation (DFG) through the International Research Training
Group (IRTG) 2101. The authors also acknowledge the German Research Foundation through the Priority Program SPP 1839 'Tailored Disorder' (PE 1524/10-2) and the German Federal Ministry of Education and Research (FKZ 13N14877, FKZ 03ZZ0434, 03Z1H534), and the European Union’s Horizon 2020 research and innovation programme under the Marie Sklodowska-Curie Grant No. 675745.

\bibliography{Mybib} 

\end{document}